\documentclass[openacc]{rstransa}

\renewcommand{\textcolor}[2]{#2}

\titlehead{Research}

\begin{document}

\title{The OASES Project: 
Exploring 
the Outer Solar System through Stellar Occultation 
with 
Amateur-Class Telescopes}

\author{
K. Arimatsu$^{1}$}

\address{$^{1}$The Hakubi center / Astronomical Observatory, Graduate School of Science,  Kyoto University \\
Kitashirakawa-oiwake-cho, Sakyo-ku, Kyoto 606-8502, Japan}

\subject{Solar System}


\corres{Ko Arimatsu\\
\email{arimatsu.ko.6x@kyoto-u.ac.jp}}

\begin{abstract}
The exploration of kilometre-sized trans-Neptunian objects (TNOs) is one of the ultimate goals in the search for the origin and evolution of the Solar System. 
However, such exploration is challenging because these small bodies are too faint to be directly detected. 
One potential avenue for detecting and investigating small and faint TNOs is the monitoring of stellar occultation events.  
This paper reviews the concept and methodology of monitoring observations of stellar occultations by small, unidentified TNOs, focusing on our observational programme, Organized Autotelescopes for Serendipitous Event Survey (OASES). 
OASES aims to detect and investigate stellar occultations by unidentified TNOs using multiple amateur-sized telescopes equipped with commercial 
Complementary Metal-Oxide-Semiconductor (CMOS)
cameras. 
Through the monitoring observations conducted so far, OASES has found one possible occultation event by a kilometre-sized TNO. 
The paper also discusses future developments of the OASES project and deliberates on the potential of movie observations in expanding the frontiers of outer Solar System research.

This article is part of the themed issue ``Major Advances in Planetary Sciences thanks to Stellar Occultations''.
\end{abstract}


\maketitle

\section{Introduction}

The investigation of trans-Neptunian objects (TNOs) 
provides fundamental knowledge of the origin and evolution of the Solar System. 
The hypothesis that TNOs are the remaining fragments of icy planetesimals renders them \textcolor{red}{an important target} for understanding the initial phase of the outer Solar System.\cite{ref1,ref2}.
For example, the size distribution of the extant Kuiper-Belt Objects (KBOs) retains the imprint of the size of the planetesimals \cite{ref3}.
The planetesimal size prior to the onset of runaway growth 
is a crucial factor in the development of planet formation theories
\cite{ref4}. The size distributions of TNOs, particularly those in the kilometre range (hereafter km-sized TNOs), which are believed to be analogous to the planetesimals, would thus provide essential insights into the accretion processes. 
Furthermore, the size distribution of the kilometre-sized TNOs 
\textcolor{red}{contains} 
clues regarding the surface evolution history of the known TNOs.
Since the surface number density of impact craters provides 
\textcolor{red}{information about} 
the surface ages of the Pluto system and (486958) Arrokoth revealed by {\it New Horizons} \cite{ref5,ref6,ref7}, determining the current size distribution and impact rate of the km-sized TNOs in the Kuiper belt and beyond are essential.

In an era of large-scale telescopic surveys, the study of small bodies in the outer Solar System remains 
\textcolor{red}{challenging}. 
The intensity of radiation from the Sun decreases inversely with the square of the heliocentric distance $D_{\rm sol}$, i.e. $\propto D_{\rm sol}^{-2}$.
and similarly the radiation from these bodies to an observer decreases with the square of their distance from the observer $D_{\rm obs}$, that is, $\propto D_{\rm obs}^{-2}$.
As a result, the apparent intensity of radiation from TNOs, which are distant from both the Sun and the Earth, decreases rapidly with distance, making comprehensive coverage of the outer regions of the Solar System, particularly at smaller sizes, extremely difficult by direct observations. 
Therefore, 
stellar occultation has become an alternative method to investigate these objects. 
By observing the transient diminution of stellar luminosity resulting from the passage of TNOs in front of them, stellar occultation allows for the measurement of size and shape with remarkable precision. 
Several programs, including the ERC Lucky Star project, have contributed to characterizing known TNOs through occultation observations, and to publishing substantial results \cite{ref_r11,ref_r12}. 
In recent years, this technique has been further developed to detect even smaller, unidentified TNOs. 
As of the mid-2010s, occultation studies focusing on kilometre- to sub-kilometre-sized TNOs had been achieved using ground-based and space-borne telescopes \cite{ref8,ref9,ref10,ref11,ref12}.
However, no program had found an occultation candidate by a km-sized TNO.

This paper presents the concept and methodology of monitoring observations of occultations by small TNOs, with a focus on the Japanese observational programme named Organised Autotelescopes for Serendipitous Event Survey (OASES). 
OASES aims to detect and investigate stellar occultations by unidentified TNOs using multiple $0.2 \-- 0.3$ metre class telescopes equipped with commercial 
Complementary Metal-Oxide-Semiconductor (CMOS)
cameras.
During the $2016 \-- 2017$ observation season, the OASES project identified a single occultation event candidate caused by a km-sized TNO for the first time.
Currently, the OASES observation system is undergoing upgrades, aiming to detect occultation events by a greater number of TNOs, including those located at the Kuiper belt and beyond.
In the following,
\S2 provides an overview of stellar occultations by distant objects.
In \S3, the results of previous OASES observations are presented.
\S4 describes the current concept and the results of performance test observations for the upgraded OASES programme.
Finally, the paper summarises with \S5, including prospects for future OASES observations.

\section{Concept of the occultation survey}

\begin{figure}[!h]
\centering\includegraphics[width=2.8in]{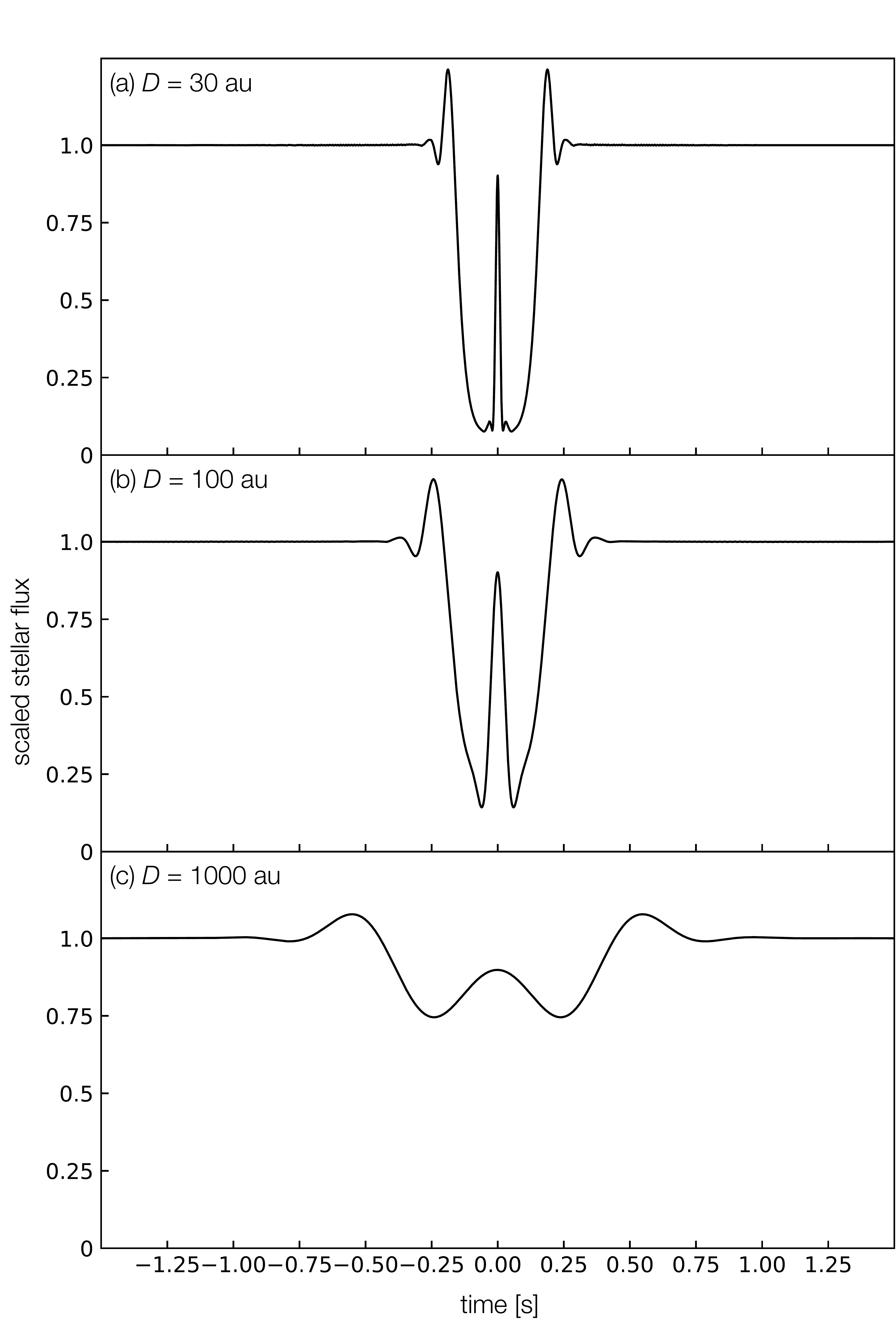}
\caption{Light curve simulation of an occultation by an r = 2.5 km- sized object located at (a) $D_{\rm obs} = 30$ au, (b) $D_{\rm obs} = 100$ au, and (c) $D_{\rm obs} = 1000$ au, respectively.
Each light curve assumes a circular shadow orbiting at a circular orbit with a semi-major axis approximately equal to its distance $D_{\rm obs}$, and passing over a background $V = 13.5 \ {\rm mag}$ A0V star with an impact factor of zero and is generated using an occultation simulator called Diffracted Occultation’s United Simulator for Highly Informative Transient Explorations (DOUSHITE) \cite{ref_r13}.}
\label{fig_LC_all}
\end{figure}

When considering the stellar occultations caused by small distant objects, it is essential to account for the effects of diffraction.
A typical scale that represents the diffraction effects is the Fresnel scale, which is defined by

\begin{align}\label{}
\begin{split}
f = \sqrt{\frac{D_{\rm obs} \lambda}{2}}  \sim 0.19 ~ \sqrt{\frac{D_{\rm obs}}{1~{\rm au}}} \ {\rm km},
\label{eq_fresnel}
\end{split}
\end{align}

where $\lambda$ is the wavelength assumed to be $\lambda = 500 ~{\rm nm}$. 
Since the spatial scale of km-sized TNOs is close to this Fresnel scale, the light curve observed during an occultation is expected to be significantly influenced by diffraction effects.

As detailed in previous studies\cite{ref13,ref14,ref101}, the diffraction effect on the light curve during an occultation is contingent upon a number of factors,  including the distance, size, and shape of the TNO body, the angular size of the occulted star, the impact parameter of the occultation path seen from the observer, the relative velocity of the TNO, and the observation wavelength.
Figure~\ref{fig_LC_all} provides examples of simulated light curves of a $V = 13.5 \ {\rm mag}$ A0V star occulted by a hypothetical distant object at opposition located at 30, 100, and 1000 au, respectively.
The light curves are generated using an occultation simulator called Diffracted Occultation’s United Simulator for Highly Informative Transient Explorations (DOUSHITE) \cite{ref_r13}.
DOUSHITE is capable of producing diffraction patterns of an arbitrary two-dimensional shadow shapes with the computational techniques proposed by ref. \cite{ref13}. 
Each light curve assumes a circular shadow of 2.5 km in radius, orbiting at a circular orbit with a semi-major axis of $\sim D_{\rm obs}$, and passing over a background star with an impact factor of zero.
It is evident that at 30 au and 100 au, stellar dimming occurs with nearly the same duration and slightly different depths. 
Unlike direct observations, where the detection limit decreases rapidly with distance, 
the effects of diffraction scale approximately with the square root of distance, i.e., $D^{1/2}_{\rm obs}$, as represented by Fresnel scale $f$ in equation~\ref{eq_fresnel}.
Consequently, the efficiency of occultation observations decreases less significantly for more distant objects compared to closer ones. This enables occultation observations to maintain sensitivity even for objects in the hypothesised "second Kuiper belt," whose existence has been suggested in recent interplanetary dust studies \cite{ref16} and direct survey observations studies \cite{ref17}.

Figure~\ref{fig_LC_all}c shows the simulated light curve assuming a circular shadow located at 1000 au. 
Although the depth of the dimming is reduced due to diffraction and the finite angular size of the star, 
detectable levels of brightness variation are still expected.
Therefore, occultation monitoring observations can detect not only small objects in the Kuiper belt region but also in the inner Oort cloud, 
which are impossible to detect through direct observations.
If observations are conducted with sufficient temporal resolution, 
allowing for the measurement of the diffraction profile, 
rough estimates of the object's distance and size can be made. 
This enables occultation monitoring observations to reveal the size-frequency distribution and spatial distribution of small objects across a wide range of space, 
from the Kuiper belt to the inner Oort cloud.

In either case, the duration of the occultation is expected to be $\sim$ one second or sub-seconds.
It is therefore necessary to gain stellar light curves with a temporal resolution of approximately 0.1 seconds or less in order to gain insight into the size and the distance of the occulting TNOs.
In addition, according to the previous occultation survey results \cite{ref11},
the event rate of the occultations by km-sized TNOs is expected to be less than $\sim 10^{-2} \, {\rm yr^{-1}}$.
Therefore, it is necessary to monitor at least $\sim 10^3$ or more stars simultaneously to detect occultation events in a reasonable observational time.
In order to ensure the robust detection of such rare and short-timescale events, 
it is favourable to implement a simultaneous monitoring strategy utilising multiple independent telescopes.

\section{The outline of the previous OASES observations}
\label{sec3}

\begin{table}[!h]
\caption{Specification of the previous and upgraded OASES observation system}
\label{tab:spec}
\begin{tabular}{lcc}
\hline
  & Original observation system & Upgraded observation system \\
  Number of systems & 2 & 4 (including 2 subsystems) \\
  Optics & \multicolumn{2}{c}{Celestron Rowe-Ackermann Schmidt astrograph \textcolor{red}{(RASA)}} \\
   \ \ Aperture  [mm]   & 279 & 279 \& 203 (subsystem)              \\
 \ \ Focal Ratio        &  f/1.58   &   f/1.58   \&  f/1.97 (subsystem)     \\
\ \ Camera             &   ZWO Co. Ltd. &  ZWO Co. Ltd.    \\
                       &   ASI1600MM-Cool    &    ASI 2600MM Pro    \\
 \  \ Number of effective pixel     & $4656\times3520$   &   $6248\times4176$  \\
 \  \  Pixel scale [arcsec ${\rm pixel}^{-1}$]    & 1.79 &  1.76 \&  TBD(subsystem)   \\
\  \ Field of view [${\rm deg.^{2}}$]  & 3.83  & 6.23 \&  TBD(subsystem)  \\
\  \ Readout noise [$e^-$]      &     2.9  &  1.5   \\ 
\  \ Peak quantum efficiency [\%]   &     $\sim 60$  & $\sim 80$ \\
 \hline 
\hline
\end{tabular}
\vspace*{-4pt}
\end{table}

The previous OASES project was initiated to detect stellar occultations caused by small TNOs, with a particular focus on km-sized KBOs\cite{ref101}. 
The project utilised two identical observation systems that offer high-cadence and wide-field imaging to detect short-timescale and rare stellar occultation events robust to false detections such as atmospheric scintillation noise.

The specification of the previous OASES observation system hardware is presented in Table~\ref{tab:spec}.
The previous system consists of a prime-focus optical tube (Celestron 11" Rowe-Ackermann Schmidt Astrograph \textcolor{red}{(RASA)}, hereafter RASA11) for amateur astronomer offering a wide field-of-view (FOV).
For the detector part, we employed a ASI1600MM-Cool CMOS camera provided by the ZWO Co. Ltd, which has a $17.7 \, {\rm mm} \times 13.4 \, {\rm mm}$ sensor area.
This effective imaging area corresponds to the $1^\circ.64 \times 1^\circ.24$ FOV.

Previous OASES monitoring observations were conducted between June 2016 and August 2017.
However, observations were only conducted on dates when the observation field was visible, weather conditions were favourable, and the observation system was in optimal condition.
In the previous observations, two observation systems were set up on Miyako Island, Miyakojima city, Okinawa Prefecture, Japan.
For monitoring, we selected an observation field at $({\rm Ra}, {\rm Dec}) = (18  {\rm h}\,  30 {\rm m} \, 00{\rm s}, -22^\circ \, 30^\prime \, 00^{\prime\prime})$. 
Given the proximity of this field to the ecliptic (ecliptic latitude $\beta \sim 0.8^\circ$) and the Galactic plane (Galactic latitude $b \sim -5.6^\circ$), 
it is anticipated that the detectability of the occultation would be enhanced.
Since Miyako Island is approximately $10^\circ$ lower latitude than the major Japanese islands,
it is more appropriate site for the observation of this low-declination field.
In order to achieve robust detections of occultation event candidates that are not affected by atmospheric scintillation effects, 
the two systems were installed at a distance of either 39 metres or 52 metres apart. 
During the observations, we monitored stars in the selected field with the exposure time of $65.0$ milliseconds.
Individual images were obtained at a frame rate of 15.4 frames per second.
At this frame rate, the OASES original system can monitor approximately 1,400 stars simultaneously\cite{ref102}. 
A total of 60 hours of imaging data was obtained with both systems during periods of favourable weather conditions, which corresponds to approximately 50 terabytes of raw data.

\begin{figure}[!h]
\centering\includegraphics[width=2.5in]{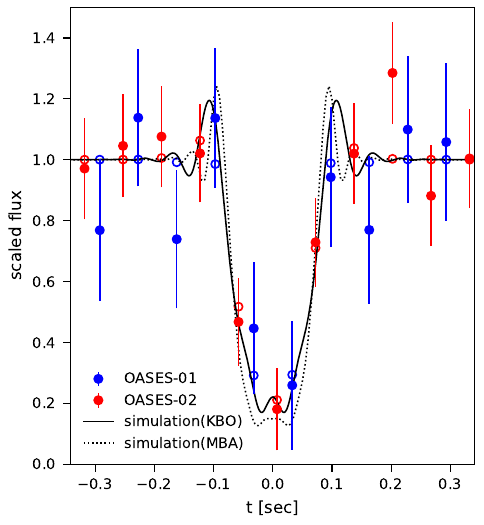}
\caption{Light curve of the candidate of the occultation event (from ref.\cite{ref102}). Blue and red filled dots with error bars are flux points of an occulted star obtained with the OASES-01 and OASES-02 systems, respectively.
$t = 0$ is set to be the central time of the occultation event candidate.
Solid curve represent the best-fit result of the theoretical light curve
(assuming an occulting object with a radius of 1.3 km at a distance of $D_{\rm obs} = 33$ au).
Open circles are the theoretical flux points obtained for each time bin.
For comparison, dotted curve is the theoretical light curve of of an occultaiton by a 500 m radius object in the main belt ($D_{\rm obs} = 2$ au).}
\label{fig_OASES_candidate}
\end{figure}

\begin{figure}[!h]
\centering\includegraphics[width=3.5in]{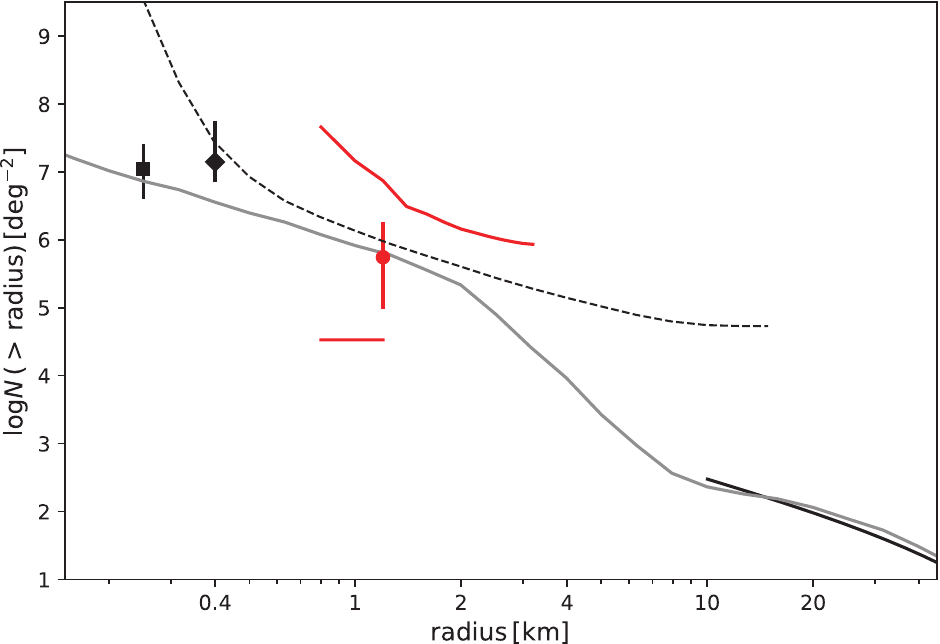}
\caption{Cumulative size distribution of KBOs in the ecliptic plane ($|\beta| < 5^\circ$), plotted as a function of radius (from ref.\cite{ref102}).
The red circle indicates the observational constraint obtained from the single detection by the OASES monitoring, 
along with a $1\sigma$ error bar. 
The upper and lower red curves are the constraints at the 95\% confidence level, respectively.
The black points are the previous results of space-borne instruments' occultation surveys, Hubble Space Telescope/Fine Guidance Sensors (square)\cite{ref10} and CoRoT (diamond)\cite{ref12}, respectively.
The 95\% upper limit obtained by the previous ground-based survey TAOS\cite{ref11} is shown as the dashed black line.
The grey line represents a theoretical collisional evolution model that assumes initial planetesimal radii of between 0.4 and 4 km\cite{ref3}.
The solid black line at radius $> 10~{\rm km}$ shows the KBO size distribution obtained by direct observations\cite{Fuentes10}.}
\label{fig_OASES_sizedistribution}
\end{figure}

The analysis of the OASES data involved the detection of simultaneous flux drops in the light curves of stars observed by the two independent systems. After processing over 26,400 time-sequential images per data run and generating 7.18 billion photometric measurements corresponding to $6.05 \times 10^4$ star hours, a single stellar occultation event candidate was identified\cite{ref102}. 
Figure~\ref{fig_OASES_candidate} shows the light curves of the occultation event candidate obtained during the previous monitoring observations.
The candidate event was captured on 28 June 2016 with the two OASES systems simultaneously.
According to the fit of the theoretical light curve to the observed light curve of the event,
this candidate event is consistent with an occultation by an object residing in the Kuiper belt with a radius of $1.3^{+0.8}_{-0.1}~{\rm km}$. 
It cannot be ruled out the possibility that the observed flux drop is the result of an occultation by a main-belt asteroid (MBA) with a radius of $\sim 500$ m (a dotted curve in Figure~\ref{fig_OASES_candidate}).
It should be noted that,
the expected number of occultations by such sub-km sized MBAs detected in the OASES monitoring is approximately $1.2 \times 10^{-3}$, according to the previous results of their surface number density revealed by Subaru/Suprime-Cam ecliptic survey\cite{ref_r14}.
Therefore, the object most likely to cause the candidate event is thought to be a KBO.
This detection represents the first observational evidence of a stellar occultation by a km-sized KBO. 
Figure~\ref{fig_OASES_sizedistribution} \textcolor{red}{presents} the observational constraint on the cumulative size distribution of KBOs in the ecliptic plane obtained from this single detection.
The number density of KBOs with radii exceeding 1.2 km was estimated to be approximately $5.5^{+12.7}_{-4.6} \times10^5 \, {\rm deg^{-2}}$ around the ecliptic.
This result supports a theoretical model of the TNO size distribution that suggests an excess population of objects with radii around 1–2 km, indicating that planetesimals in the primordial outer Solar System would likely have grown into km-sized bodies before undergoing runaway growth. 

\section{Current status of the upgraded OASES program}
The rapid progress of performance of CMOS sensors in consumer products, 
along with the increased capacity and accelerated data transfer speeds of storage for personal computers at commercial levels, 
continue to improve even after the previous OASES project. 
This enables the realisation of more efficient occultation monitoring programs.
The primary objectives of the upgraded OASES program are as follows:
\begin{description}
    \item[More robust detection of km-sized KBOs.] 
    Since the previous OASES campaign achieved the detection of the occultation event candidate by a km-sized TNO for the first time, its results and discussions are still being debated\cite{Nir24}. 
    For example, as already described in \S~\ref{sec3},
    the possibility of a false detection due to a main-belt asteroid occultation cannot be entirely ruled out.
    The upgraded campaign thus aims to achieve multiple detections of the occultation events by KBOs and obtain more robust observational constraints of their cumulative number densities.
    Assuming the constraints of the KBO size distribution obtained by the previous study \cite{ref101}, 
    the expected number of occultation events by KBOs over the five-year monitoring period, using the upgraded systems, is within the range of $10^1\-- 10^2$.
    \item[Exploration of "second Kuiper belt".]
    Recent in-situ studies by {\it New Horizons} reported overabundance of interplanetary dust particles at the outer edge of the Kuiper belt, which is a possible evidence for another ecliptic population outside the known Solar System\cite{ref16}. 
    The presence of the outer population is also suggested by recent deep surveys conducted by Subaru/Hyper Suprime-Cam\cite{ref17}. 
    However, direct observations suffer from significant observational biases, making it challenging to obtain its radial extent and total number of objects. 
    Observations of occultation events are expected to make invaluable contributions to understanding the characteristics of this new population of objects.
    \item[Occultation evidence for Oort cloud.]
    The ultimate goal of the OASES program is to detect stellar occultation events by objects in the Oort cloud.
    Since the Oort cloud objects are extremely faint, making direct detection challenging,  monitoring stellar occultations remains the most viable observational methods.  
\end{description}

To achieve the objectives, it is necessary to simultaneously observe a greater number of stars and detect a larger number of occultation event candidates. 
We are thus upgrading the OASES observation system.
In the following, we present current basic specifications of the commissioning upgraded OASES observation systems and the results of their performance test observations in 2024.

\subsection{Basic specification}
\begin{figure}[!h]
\centering\includegraphics[width=3.5in]{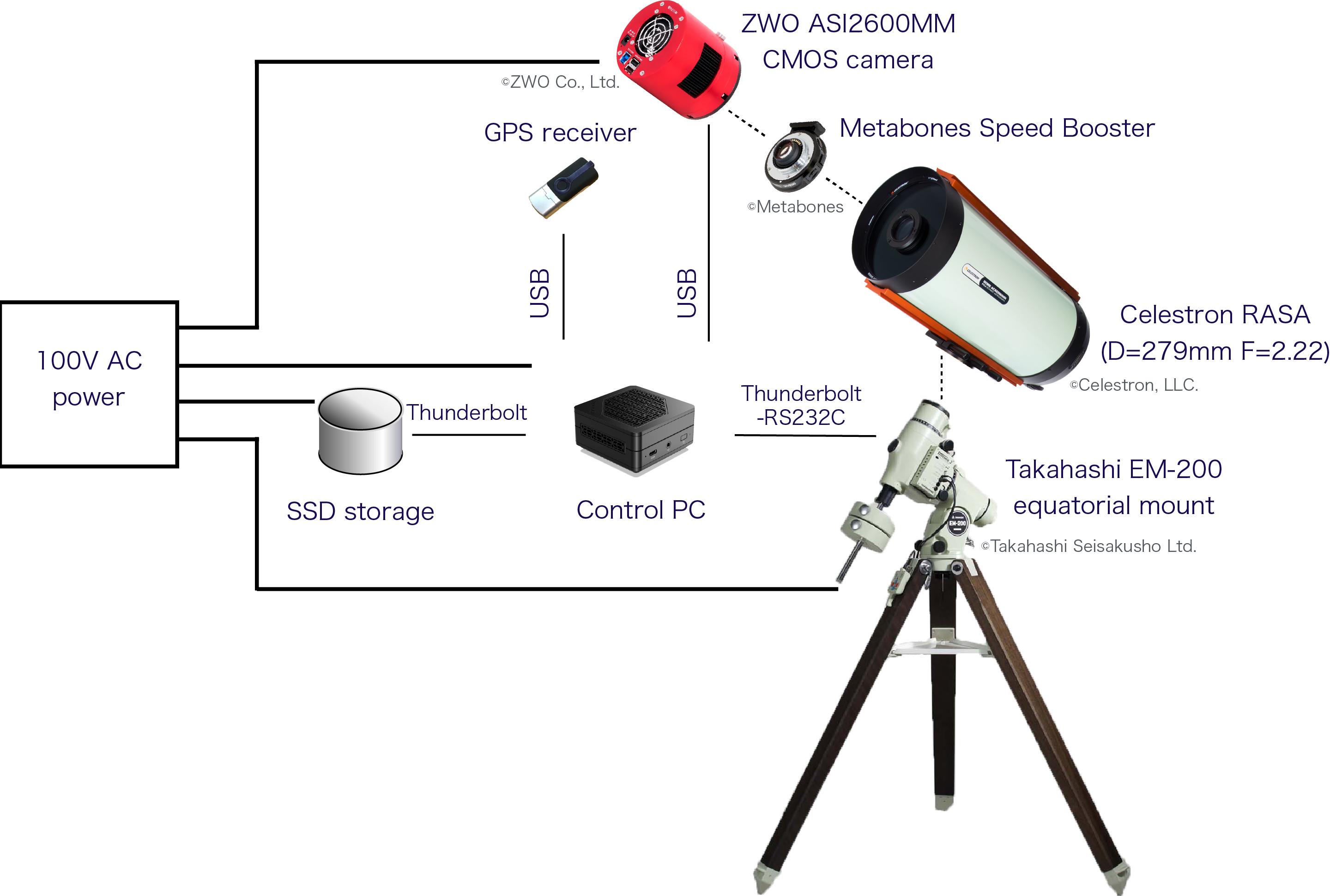}
\caption{
Overview of the functionality of the OASES upgraded main observation system. 
The ASI2600MM Pro CMOS camera is affixed to the RASA11 optical tube via a focal reducer. 
The single control PC is responsible for the operation of both the CMOS camera and the mount, with communication facilitated by USB links.  
A 100 V AC power supply is distributed to all electrical devices.}
\label{fig_OASES_systemchart}
\end{figure}

The specification of the upgraded observation systems is presented in Table~\ref{tab:spec}, 
and the overview of the functionality of the main system is shown in Figure~\ref{fig_OASES_systemchart}.
The upgraded system continues to employ the RASA11 optical tube, as used in the previous system.
The primary difference between the original and the upgraded systems is the ZWO Co. Ltd. ASI2600MM Pro CMOS camera.
To monitor as many stars as possible,  
larger format ($23.5 \, {\rm mm} \times 15.7 \, {\rm mm}$ sensor area offering $3^\circ.05 \times 2^\circ.04$ FOV) and low readout noise ($\sim 1.5 \, e^-$) CMOS sensors are adopted for the upgraded systems.
These sensors enable us to gain wider fields and higher-sensitivity sequential images of the observation field.
In addition, the additional subsystems each consisting of a 203~mm aperture RASA optical tube are planned to be installed to achieve more robust detections of occultation events.  
These systems are planned to be installed with separations of approximately $50 \-- 500$~meters.

\subsection{Performance test observations}

\begin{figure}[!h]
\centering\includegraphics[width=3.5in]{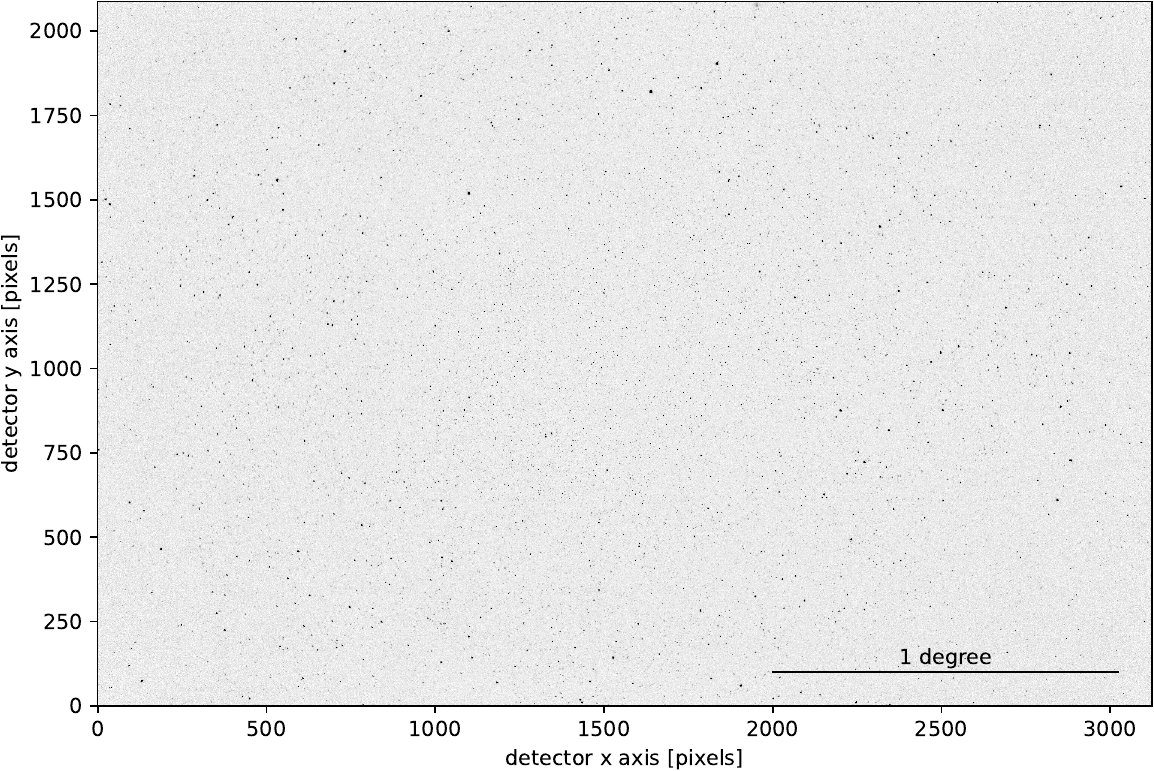}
\caption{Example of the image of the selected field $({\rm Ra}, {\rm Dec}) = (18  {\rm h}\,  30 {\rm m} \, 00{\rm s}, -22^\circ \, 30^\prime \, 00^{\prime\prime})$ obtained with the OASES performance test observations. }
\label{fig_img_testdata}
\end{figure}

In July 2024, we carried out OASES performance test observations at the Ishigakijima Observatory on Ishigaki Island, Okinawa, an observation site candidate of the upgraded program. 
The single upgraded system was placed on the rooftop of the observatory. 
For the test observations, 
we selected an observation field at $({\rm Ra}, {\rm Dec}) = (18  {\rm h}\,  30 {\rm m} \, 00{\rm s}, -22^\circ \, 30^\prime \, 00^{\prime\prime})$, following the previous monitoring\cite{ref102}.
Sequential images of the field were obtained with an exposure time of 144 milliseconds, which corresponds to the frame rate of $\sim 6.95$ fps. 
The acquired images were subjected to processing via the OASES data reduction pipeline\cite{ref101}, encompassing dark subtraction through the utilisation of dark frames, flat-fielding, and the sky background subtraction. 
An exemple of the reduced image obtained subsequent to these procedures is illustrated in Figure~\ref{fig_img_testdata}.

\begin{figure}[!h]
\centering\includegraphics[width=2.5in]{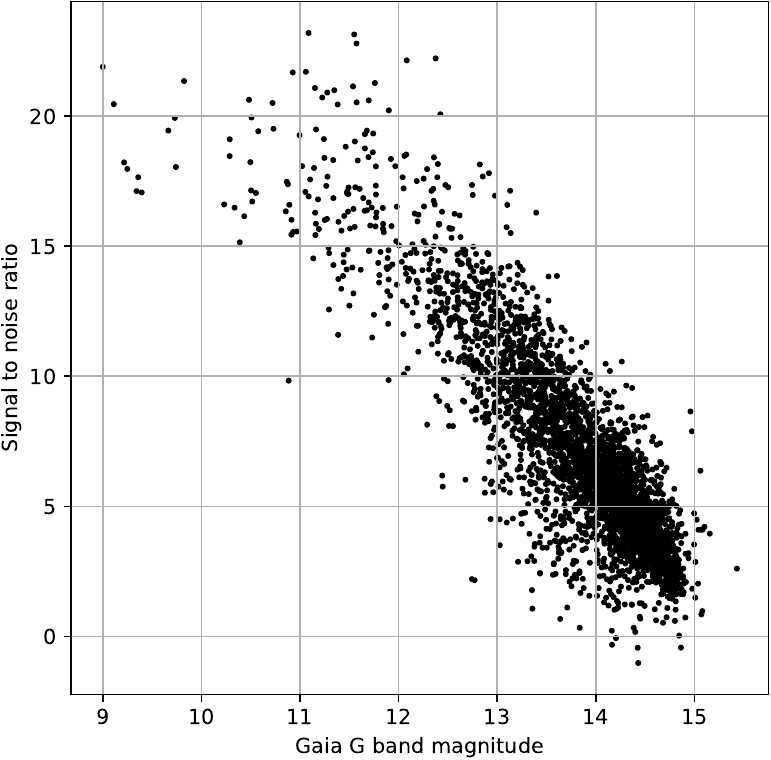}
\caption{Example of LCSNR values for stars close to the centre of the observation field $({\rm Ra}, {\rm Dec}) = (18  {\rm h}\,  30 {\rm m} \, 00{\rm s}, -22^\circ \, 30^\prime \, 00^{\prime\prime})$ as a function of Gaia G band magnitude $m_{\rm G}$ obtained in the OASES performance test observations.}
\label{fig_SN_hist}
\end{figure}

\begin{figure}[!h]
\centering\includegraphics[width=2.5in]{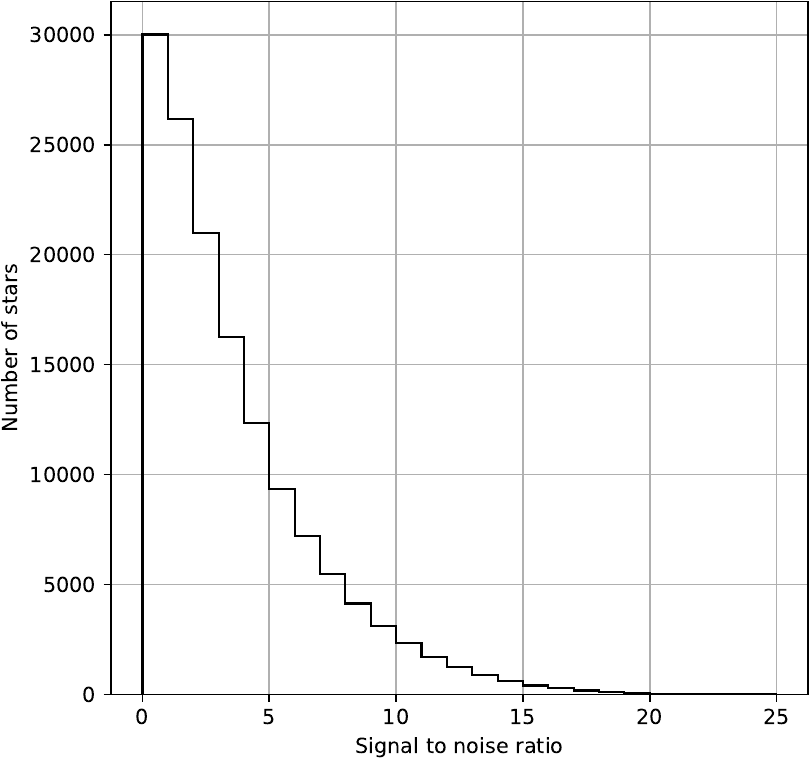}
\caption{ 
\textcolor{red}{Cumulative} distribution of LCSNR for the stars in the selected field $({\rm Ra}, {\rm Dec}) = (18  {\rm h}\,  30 {\rm m} \, 00{\rm s}, -22^\circ \, 30^\prime \, 00^{\prime\prime})$ relative to the LCSNR observed in the OASES performance test observations.}
\label{fig_SN_Gband}
\end{figure}

For each star in the image with the Gaia \textcolor{red}{G-band magnitude $m_{\rm G}$ of brighter than $\sim$15}, we perform aperture photometry following the previous OASES data reduction.
Following the photometric analysis,
the mean intensity for each light curve and its standard deviation are calculated in order to determine the signal-to-noise ratio of the light curve, which is referred to as LCSNR\cite{ref101}.
Figure~\ref{fig_SN_Gband} shows an example of the LCSNR values relative to 
$m_G$ for stars located close to the field centre obtained from the test observation data.
The average $m_{\rm G}$ for which LCSNR = 5, 10, and 15 are $m_{\rm G} \sim 14.3$, 13.3, and 12.2, respectively.
However, it should be noted that
the LCSNR for stars located near the edges of the image gradually decreases due to vignetting.
Figure~\ref{fig_SN_hist} illustrates a 
\textcolor{red}{cumulative} 
distribution of the LCSNR of the stars in the selected observation field that were observed with the upgraded system.  
The observation system has the capacity to monitor approximately $\sim 9,300$, 2,300, and 400 stars with LCSNR values greater than 5, 10, and 15, respectively.
The performance test observations proved that the OASES system can simultaneously monitor up to $\sim$ 10,000 stars. 

\section{Summary and prospects}

This paper has presented the concept and methodology of the OASES project, which aims to detect stellar occultations caused by small TNOs using amateur-class telescopes. 
The previous observations conducted between 2016 and 2017 successfully identified a candidate occultation event by a km-sized TNO, representing the first such detection. 
The analysis of this event suggests that the OASES approach can provide valuable insights into the size distribution of small TNOs, which are challenging to detect via direct observations.

The upgraded OASES program aims to improve the detection capabilities of the system, 
allowing for the monitoring of a larger number of stars and increasing the likelihood of multiple TNO occultation detections. 
The improvements of the upgraded OASES systems will enable more robust and efficient occultation monitoring, 
with the potential to explore more distant regions, 
including the hypothesised "second Kuiper belt" and the inner Oort cloud. 

Currently, we are preparing for continuous monitoring observations by expanding the upgraded observation systems and refining our data analysis techniques.
During the test observations, the frame rate was constrained to 7 fps due to the limitations of the control PC setup.
However, in the future, we plan to implement observation modes with higher frame rates, 
allowing us to capture brightness variations in even shorter timescales.
Data analysis speed and data storage capacity are currently the main challenges to be overcome in this project.
The current data reduction pipeline is a prototype developed for processing test observation data, which has limitations in processing speed and storage capacity. 
The pipeline can process approximately 2.5 terabytes of data, which is equivalent to that obtained in an hour of observation, in approximately 7.7 hours. 
This processing time is marginally acceptable because observation downtimes due to bad weather, daylight hours, and planned observational breaks during certain seasons provide additional time for data processing.
Nevertheless, further improvements in speed are necessary to establish a more robust data analysis system.
Additionally, the present data storage for the test observations is managed with a 24-terabyte system. 
We therefore plan to speed up the pipeline and develop a computing system
with an increased storage capacity of around 1200 terabytes.

In addition to detecting occultations by unknown TNOs, 
as introduced in this paper, 
the OASES project will also conduct campaign observations for occultation events by larger, known TNOs\cite{Arimatsu19b,Arimatsu20}. 
Furthermore, we plan to search for extremely short-timescale events on planetary surfaces, such as Jupiter impact flashes\cite{Arimatsu22}, as well as unknown transients in deep space\cite{Arimatsu21}. 
Through the observation of these diverse phenomena, the OASES project aims to contribute to the emerging field of "movie astronomy," 
where continuous monitoring and high-frame-rate observations open new windows into the dynamic phenomena occurring in the Solar System and beyond.

\ack{
\textcolor{red}{
We thank the anonymous referees for their careful reviews and for providing constructive suggestions.}
This research has been made possible thanks to the generous support of the Japan Society for the Promotion of Science (JSPS) KAKENHI Grant (18K13606, 21H01153).}


\end{document}